\documentclass[runningheads]{llncs}
\usepackage[T1]{fontenc}
\usepackage{graphicx,verbatim}
\usepackage{amsmath}
\usepackage{caption}
\begin{document}
\title{An Interactive Mobile Augmented Reality System for Efficient Virtual Model-to-Patient Registration in Thoracic Surgery}
%

\author{Emmanuel Ajaero\inst{1} \and
Pascal Spiegler\inst{1,3}\and
George Rakovich\inst{2}\ \and \\
Marta Kersten-Oertal\inst{1}\and
Yiming Xiao\inst{1}\thanks{Corresponding author.}}

\authorrunning{Emmanuel Ajaero et al.}

\institute{Department of Computer Science and Software Engineering, Concordia University, Montreal, Canada \\ 
\and
Maisonneuve Rosemont Hospital, University of Montreal, Montreal, Canada
\\
\and
Faculty of Medicine, McGill University, Montreal, Canada
\\
    \email{yiming.xiao@concordia.ca}}
  
\maketitle              
\begin{abstract}
Video-Assisted Thoracoscopic Surgery (VATS) is a minimally invasive procedure, where precise surgical port placement is vital, yet it relies heavily on surgeon experience and mental mapping of pre-operative imaging in the operating room. To address this challenge, we present an accessible, interactive mobile augmented reality (AR) framework, \texttt{HILMARTS}, to enable accurate and efficient pre-operative virtual model-to-patient registration for AR-guided surgical navigation. Our system runs on a commercially available LiDAR-equipped iPad and supports a user-in-the-loop workflow through a two-stage coarse-to-fine registration process. First, the user performs a manual alignment using the AR view on the iPad. This initial maneuver is then refined through automatic point-cloud registration leveraging the LiDAR sensor. We further introduce a novel occupancy-grid-based \textit{Surface Coverage Score} to provide real-time, actionable feedback to the operator on the geometric completeness of LiDAR data capture for reliable point-cloud-based alignment. In a phantom study with 16 participants, the system achieved a 3.31$\pm$1.26mm registration accuracy in $\sim$4 min with a mean System Usability Scale (SUS) of 84.1. These findings highlight the system’s accuracy, efficiency, robustness, and potential clinical usability, and could support its applicability to other AR-based surgical navigation tasks.


\end{abstract}
\section{Introduction}
Video-Assisted Thoracoscopic Surgery (VATS) has become a standard of care for many thoracic procedures due to its minimally invasive nature, which is associated with reduced post-operative pain and shorter hospital stays~\cite{yan2014video,sadeghi2022current,drevet2016uniportal}. A key determinant of a successful VATS procedure is the precise placement of surgical ports, which directly impacts instrument access, visualization, and overall operative efficiency~\cite{yan2014video}. Currently, this planning step relies predominantly on the surgeon's experience and their ability to mentally correlate pre-operative CT or MRI scans with the patient's anatomy — a process that is inherently subjective and variable. 

Augmented Reality (AR) technologies offer an opportunity to address this issue by overlaying a patient-specific digital model along with surgical plans \cite{harirpoush2024virtual}, derived from pre-operative imaging, directly onto the surgical field~\cite{sadeghi2022current,bakhuis2023essential}. The fundamental prerequisite for any such system is an accurate registration between the patient-specific digital model and the patient. However, this registration problem is non-trivial. While head-mounted displays (HMDs) like the Microsoft HoloLens have been explored, their built-in sensors often lack the resolution needed for precise alignment~\cite{gsaxner2023hololens}. Consequently, existing solutions frequently rely on specialized, expensive external optical tracking systems, limiting accessibility and practical deployment in diverse clinical environments. By contrast, a handheld tablet offers an easy-to-sterilize, accessible, and flexible AR tool with a minimal clinical footprint. While AR tracking is relatively mature in neurosurgery, abdominal procedures, and orthopedics~\cite{fick2021current,jud2020applicability}, AR-guided thoracic surgery~\cite{sadeghi2022current} is relatively under-explored. Notably, many existing works demonstrated the feasibility of overlaying 3D anatomical models, but often required cumbersome fiducial/marker-based registration and/or relied heavily on HMDs, which can be ergonomically challenging for surgeons, and require a significant investment~\cite{gsaxner2023hololens} and dedicated technical support. Alternatively, recent markerless registration methods overcome the coarse spatial mapping of standard AR headsets~\cite{hu2020alignment,nguyen2020augmented} by using complex external hardware, such as dedicated 3D scanners, tracking arrays, or hybrid C-arms~\cite{glas2021augmented,fick2021holographic,condino2021hybrid,li2020smartphone}, which can still be costly. 

While built-in Time-of-Flight sensors enable direct surface acquisition and point cloud registration~\cite{gsaxner2021augmented,gu2021feasibility}, challenges for robust and efficient workflows remain. The Iterative Closest Point (ICP) algorithm, particularly the Point-to-Plane variant~\cite{chen1991object}, is widely used for fine alignment, but requires a sufficiently accurate initialization to avoid convergence to local minima~\cite{besl1992method}. Global registration methods, such as those using Fast Point Feature Histograms (FPFH)~\cite{rusu2009fast} with RANSAC-based~\cite{fischler1981random} correspondence search~\cite{zhou2018open3d}, provide robust initial alignments. However, they are highly sensitive to the presence of non-target objects, such as the operating table, equipment, and medical personnel. Recent advances in real-time horizontal pose estimation, such as the BlazePose model~\cite{bazarevsky2020blazepose}, which provides 33 body keypoints and a per-pixel segmentation mask, offer a promising solution to semantically isolate the relevant anatomy (i.e., the patient's torso) from raw depth/surface sensor data prior to geometric registration.

In this work, we present a novel end-to-end mobile AR framework for virtual anatomical model-to-patient registration designed for thoracic surgery with the potential to extend to other surgical applications, called \texttt{HILMARTS} (Human-In-the-Loop Mobile Augmented Reality for Thoracic Surgery). \textit{Our system is distinguished by three key contributions}: \textbf{First}, we propose an accessible mobile-AR-based strategy that uses the tablet device as a maneuver to interactively align virtual model with patients manually. \textbf{Second}, we introduce an efficient and robust point-cloud-based registration utilizing the tablet's  LiDAR sensor to refine the registration \textit{without external tracking}. \textbf{Lastly}, we design a task-specific user interface incorporating an intuitive feedback score, enabled by a novel projection-based algorithm, to assess LiDAR data quality and ensure reliable registration refinement. We demonstrate the usability, efficiency, robustness, and accuracy of the proposed system through a user study of 16 participants.

\section{Methods and Materials}

\subsection{System Overview}

\begin{figure}[ht]
\includegraphics[width=\textwidth]{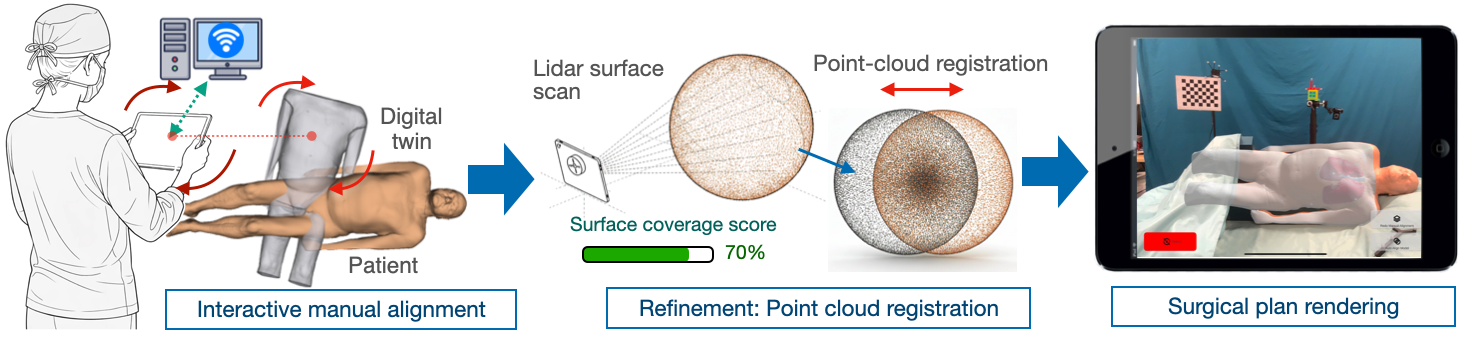}
\caption{Demonstration of the workflow of the proposed AR system.} \label{fig1}
\end{figure}

We propose a mobile AR framework for patient-specific model–to–patient registration in thoracic surgery. As illustrated in Fig.~\ref{fig1}, the system adopts a two-stage coarse-to-fine workflow: (1) interactive manual alignment of the patient-specific model in AR, followed by (2) automatic point-cloud-based refinement using LiDAR-acquired patient surface data. The \emph{Unity} client runs on a LiDAR-equipped iPad (Pro 2020 or later), handling AR visualization, user interaction, and synchronized RGB–depth capture via ARKit, while compute-heavy point-cloud registration runs on a local server over a secure wireless link to preserve in-clinic privacy.

\subsection{Digital and Physical Phantom Construction}
To evaluate the proposed system, we constructed a simulated thoracic surgical setup with the patient positioned laterally on a draped operating table. A patient model was generated from the Visible Human Project (Male) CT dataset (1×1×1 mm³ resolution)~\cite{mccarville2023visible}. The external body surface was segmented in 3D Slicer~\cite{fedorov20123dslicer} to produce a 3D surface mesh. The resulting mesh, together with the segmented organs, served as the virtual model and was imported into Unity as a standard 3D model asset, with the mesh forming the source point cloud. A full-scale physical phantom was fabricated using a Modix BIG-60 3D printer with PLA filament based on the CT-derived geometry. The surface was coated with matte paint to reduce specular reflection and approximate tissue appearance under clinical lighting. Eighteen landmarks (Fig.~2A) were placed on the anterior (camera-facing) surface of the phantom with pairwise correspondence to the virtual model for registration accuracy assessment.

\subsection{Interactive Manual Coarse Alignment}
To initialize registration, we introduce a device-coupled alignment strategy that leverages the tablet's natural 6-DoF motion to provide an intuitive global pose initialization of the virtual model. After loading the virtual model into the AR scene, the user activates a ``grab'' mode with a prolonged button press, causing the virtual model to mirror the incremental motion of the camera effectively coupling it to the physical device. Rather than adjusting on-screen transformation parameters, the user performs alignment through natural physical movement of the tablet while visually assessing alignment quality in AR. This interaction yields a reliable initial pose estimate and substantially reduces the search space for subsequent refinement.

\begin{figure}[t]
\includegraphics[width=\textwidth]{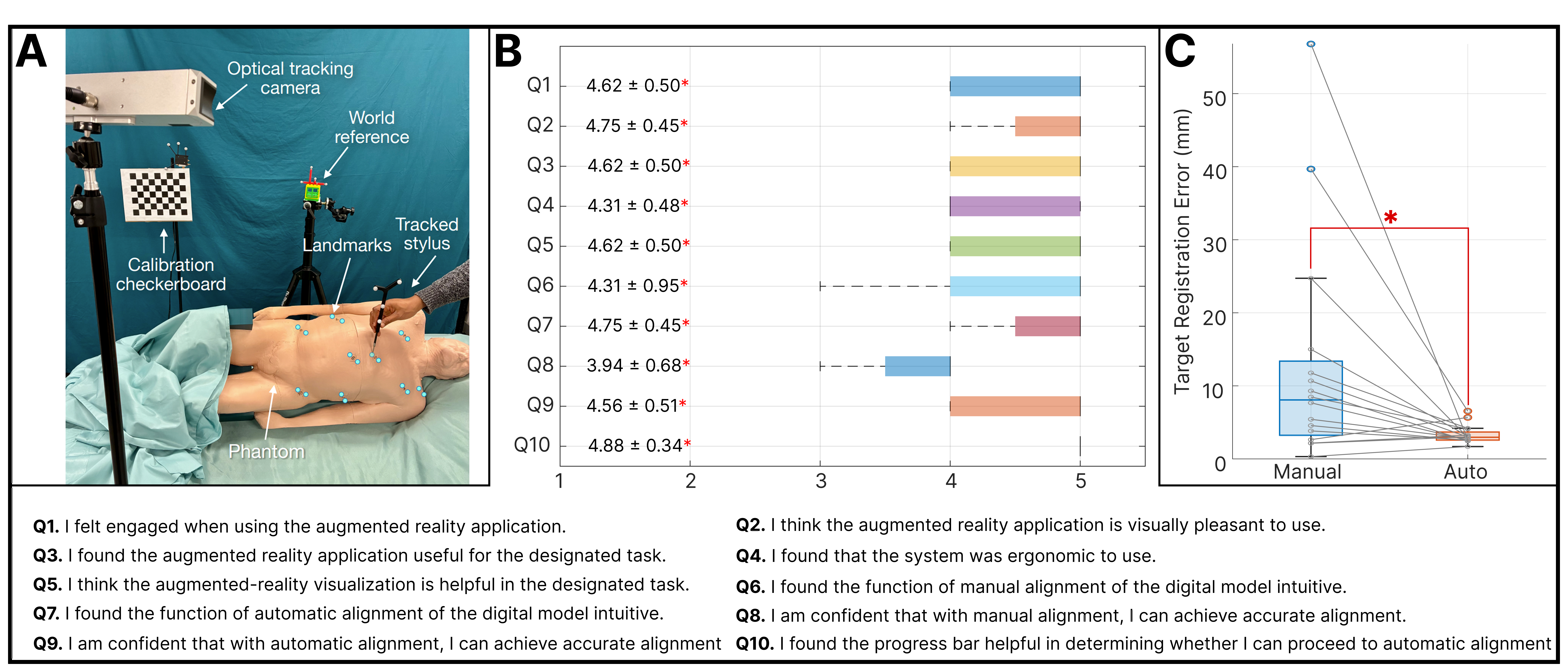}
\caption{A. Experimental setup for system validation; B. Boxplots of UX design questionnaire; C. Mean target registration errors (mTREs) from all participants between using manual registration and point-cloud-based automatic refinement (the corresponding data points between Manual vs. Auto registration are linked with a line). The * sign indicates a statistically significant pair-wise difference (p<0.05). } \label{fig2}
\end{figure}

\subsection{Surface Acquisition and Registration Pipeline}
\noindent \textbf{Surface mesh acquisition \& processing:} Following coarse alignment, the user positions the iPad to capture synchronized RGB images, LiDAR depth maps, camera intrinsics, and 6-DoF camera pose via \textit{ARKit}. These data are used to reconstruct a patient-specific 3D point cloud for registration. Depth processing is performed in two stages. First, invalid measurements are removed using physical range constraints and statistical outlier rejection, and edge-preserving bilateral filtering is used to reduce sensor noise.
Second, anatomically relevant regions are isolated from the depth map. We apply \textit{BlazePose}~\cite{bazarevsky2020blazepose} to the corresponding RGB image to extract torso landmarks and generate a segmentation mask. This mask is then projected onto the processed depth map to retain only 3D points belonging to the torso region. Finally, density-based clustering (DBSCAN)~\cite{ester1996density} is applied to suppress residual sparse outliers and improve point cloud quality prior to registration.

\noindent \textbf{Surface coverage score:} To ensure point-cloud-based registration quality, the trimmed 3D points should sufficiently cover the body surface. We introduce the \textit{Surface Coverage Score} as real-time feedback to indicate when accumulated image captures achieve sufficient surface coverage (i.e., 70\%) to trigger registration. First, an oriented bounding box for the patient is estimated via PCA, and 3D points are projected onto its dominant plane that is discretized into a $G \times G$ occupancy grid, and connected empty components are identified. Let $\mathcal{I}_{max}$ denote the largest connected empty component. The final score is defined as $\text{Coverage} =
\left(
1 - \frac{|\mathcal{I}_{max}|}{G^2}
\right) \times 100\%$. By penalizing large contiguous gaps, this metric gives a real-time completeness estimate, and registration is triggered only when coverage exceeds the threshold, mitigating under-constrained registration.

\noindent
\textbf{Point cloud registration:} Point clouds of the patient and virtual model are aligned hierarchically. They are first downsampled and represented using Fast Point Feature Histograms~\cite{rusu2009fast}. RANSAC-based correspondence search~\cite{fischler1981random} estimates candidate rigid transformations via three-point sampling with SVD-based pose estimation, yielding an initial alignment.
This initialization is then refined using point-to-plane iterative closest point algorithm~\cite{chen1991object}. The final rigid transformation is applied to the virtual model for AR visualization.

\section{User Study Design and Evaluation Metrics}

We conducted a controlled user study to evaluate the performance of the system. In all trials, participants first performed manual coarse alignment, followed by a structured surface point-capture stage. During point capture, users were instructed to acquire surface scans from the camera-facing side of the phantom prior to automatic registration. To assess the effectiveness of the \textit{Surface Coverage Score} feedback, the participants completed the registration tasks in two conditions, with the first using the guidance of the score for point capture and the other with the score hidden.

\subsection{Participant Recruitment}
Upon informed consent, 16 participants (age= 27.13 $\pm$ 5.26 yo, 6 females \& 10 males) with 1 clinician and 15 STEM students were recruited for the study. To understand potential background effects on system interaction, participants self-reported their familiarity with AR and human anatomy using a 5-point Likert scale (1 = not familiar at all, 5 = highly familiar). Mean familiarity scores were 4.07 $\pm$ 1.34 for AR and 3.40 $\pm$ 1.64 for human anatomy, indicating moderate-to-high prior exposure across the cohort.

\subsection{Usability and Experience Assessment}
Perceived usability was evaluated using the System Usability Scale (SUS)~\cite{brooke1996sus}. The SUS consists of 10 items rated on a 5-point Likert scale and is designed to assess perceived usability, including system complexity, ease of interaction, and user confidence. Responses are scored using a standardized procedure: for positively worded (odd-numbered) items, one point is subtracted from the response value, whereas for negatively worded (even-numbered) items, the response is subtracted from five. The adjusted item scores are then summed and multiplied by 2.5, yielding a maximum score of 100. According to established benchmarks, a score above 68 is generally interpreted as indicating acceptable to good usability \cite{brooke2013sus}. In addition to the SUS, a custom user experience questionnaire was administered to assess specific factors not captured by general usability metrics. It consisted of 10 Likert-scale (1-5) items evaluating perceived alignment confidence, intuitiveness of manual and automatic alignment, visual quality, ergonomics, and perceived usefulness of the \textit{Surface Coverage Score} (see Fig. 2). 

\subsection{Registration Accuracy and Efficiency Evaluation}

Registration accuracy was quantified using Target Registration Error (TRE) measured across the 18 asymmetrically distributed landmarks. Ground-truth landmark positions were acquired in the world reference frame using an optical tracking system (OptiTrak Duo) and a tracked probe. TRE was computed as the Euclidean distance between each ground-truth landmark and its corresponding transformed virtual counterpart. The mean TRE (mTRE) across all landmarks was used to quantify both manual coarse alignment error and the final residual error after automatic point-cloud-based refinement for each trial. Efficiency was assessed by recording total task completion times.

\subsection{Evaluation Data Analysis}

 Manual and automatic TRE values, as well as custom questionnaire responses (5-point Likert scale), violated normality assumptions (Lilliefors test) and were therefore analyzed using non-parametric tests. Specifically, manual coarse alignment mTRE was compared with automatic registration mTRE within trials using paired Wilcoxon signed-rank tests. Custom questionnaire items were evaluated using one-sample Wilcoxon signed-rank tests against the neutral midpoint (3). SUS scores satisfied normality assumptions and were analyzed using a one-sample t-test against the benchmark value of 68. Statistical significance was set at p < 0.05. To complement the quantitative evaluations, a qualitative analysis was conducted to capture user perspectives that may not be reflected in structured rating scales. Participants were asked two open-ended questions: (1) “What are the positives and negatives of the system?” and (2) “Do you have any suggestions for improvement?”. These freeform responses were then carefully reviewed, with key words summarized based on frequencies of occurrence across participants, providing additional context for other measures and areas for future improvement.

\section{Results}

\subsection{Registration Accuracy \& Task Efficiency}
\textbf{Accuracy:} Initial manual coarse alignment yielded a mean TRE of 12.82 $\pm$ 15.40 mm across all 16 trials (Fig.~\ref{fig2}C). The substantial variance was primarily driven by difficulty in visually estimating depth, particularly among users with limited experience in AR and human anatomy. Following automatic point-cloud-based refinement, the mTRE significantly decreased to 3.31 $\pm$ 1.26 mm (p=0.0052). This substantial reduction, alongside a drastically tightened standard deviation, confirms the necessity and efficacy of the automatic registration stage in achieving consistent surgical precision regardless of the initial manual alignment quality. When comparing interface conditions, automatic registration with the real-time \textit{Surface Coverage Score} feedback resulted in a lower mean error (3.31 $\pm$ 1.26 mm) compared to the no-feedback condition (3.99 $\pm$ 2.50 mm). Although not statistically significant (p = 0.33), the ``with feedback" condition showed reduced variability, suggesting the coverage metric improves robustness and stability rather than mean accuracy.

\noindent
\textbf{Efficiency:} Participants completed the manual coarse alignment rapidly, requiring an average of 87.55 $\pm$ 59.05 seconds. The subsequent structured point cloud collection stage took an average of 149.36 $\pm$ 124.93 seconds. During this phase, participants occasionally had to restart the capture process if their initial scan failed to exceed the required 70\% Surface Coverage threshold. However, restarts were highly infrequent, averaging only 0.25 $\pm$ 0.45 resets per user (specifically, only 4 out of the 16 users required a single point capture reset). Overall, this time investment directly translated to the significant improvements observed in registration accuracy, demonstrating that markerless registration with the low-quality LiDAR scanner on the iPad is both feasible and highly effective for an intraoperative workflow.

\subsection{Usability and User Experience}
\textbf{Quantitative data:} Our system demonstrated an excellent level of perceived usability, achieving a mean SUS score of 84.06 $\pm$ 10.82. This score is significantly higher than the established acceptable usability benchmark of 68 ($p < 0.001$), indicating that the system is highly accessible even to users with varying levels of AR and anatomical familiarity. Results from the custom 10-item user experience questionnaire (Fig.~\ref{fig2}B) further demonstrated consistently positive perceptions across all evaluated dimensions. All items were rated significantly above the neutral midpoint of 3 ($p < 0.001$ for all items). Mean ratings ranged from 3.94 $\pm$ 0.68 to 4.88 $\pm$ 0.34, reflecting strong agreement regarding interaction intuitiveness, alignment confidence, visual clarity, and perceived clinical relevance. Importantly, items specifically related to the \textit{Surface Coverage Score} and feedback-guided point capture received high ratings (Q10: 4.88 $\pm$ 0.34), indicating that participants found the real-time geometric completeness feedback highly useful. 

\noindent
\textbf{Qualitative data:} Fourteen of sixteen participants provided written feedback. Responses were largely positive and consistent with the high SUS and UX scores. \textit{Ease of use} was the most frequently cited strength, with \textit{8/14 participants} describing the system as easy, intuitive, or enjoyable. \textit{5/14 participants} emphasized its clinical relevance, noting that it addresses an important problem and could support thoracic procedures (e.g., providing an ``Xray view''). In contrast, \textit{3/14 participants} reported that manual coarse alignment was occasionally challenging, particularly for depth and rotation control. Overall, the qualitative findings support the system’s usability and perceived clinical utility.

\section{Discussion}
We present a robust, fast, markerless mobile AR framework for virtual model-to-patient registration in thoracic surgical navigation without external tracking, combining intuitive device-coupled manual alignment, segmentation-based surface capture, and real-time feedback-guided point collection. 

Our user study confirms the efficacy of the two-stage coarse-to-fine registration approach, with the automatic refinement further enhancing the alignment quality while the manual alignment reduces the search space of the point-cloud registration solution. Additionally, the final 3.31 mm accuracy is within the surgical standard of thoracic procedures \cite{zhou2020intraoperative}. We will further improve the accuracy through enhanced registration algorithms in the future. With a high SUS score of 84.1, minor ergonomic concerns (e.g., tablet fatigue) reported by some users can be mitigated by short usage time. Note that our single-trial design evaluates ``walk-up-and-use" intuitiveness for first-time users while minimizing learning effects~\cite{schmettow2017}, focusing on usability, workflow integration, and the contribution of the refinement stage. 

The proposed system still has a few limitations. First, because the CT was acquired supine, lateral positioning may induce tissue deformation affecting accuracy; we plan to address this with deformable models. Second, the feedback and no-feedback conditions were presented in a fixed order so participants could first experience the full workflow, which may introduce ordering effects; we therefore treat the standalone impact of the Surface Coverage Score cautiously and leave randomized multi-session evaluation to future work. Finally, while this work was completed in close collaboration with an experienced thoracic surgeon, the evaluation cohort primarily comprised STEM trainees with only a single active clinician. Although evaluation with first-time non-experts remains highly informative for assessing baseline intuitiveness and acceptability at this preliminary stage~\cite{chaniaud2020}, it does not fully represent the ultimate user base. Future work will focus on clinician-centered evaluations and broader adoption frameworks.

\section{Conclusion}
We have proposed a novel mobile AR platform \texttt{HILMARTS} to allow markerless virtual model-patient registration for AR-guided thoracic surgical navigation. Through preliminary evaluation, the system demonstrated high accuracy, robustness, and usability. We will seek clinical validation and extensions in other surgical applications of the \texttt{HILMARTS} system in the future.

\subsubsection*{\discintname}
The authors have no competing interests to declare that are relevant to the content of this article.




%

%
%

\bibliography{sn-bibliography.bib}
\bibliographystyle{splncs04}
%




\end{document}